\documentclass[10pt,pra,aps,showpacs,twocolumn,amsmath,amssymb,superscriptaddress,longbibliography,nofootinbib]{revtex4-1}
\usepackage{times}
\usepackage[dvips]{color}
\usepackage{graphicx} 
\usepackage{bm}

\usepackage{color}

\usepackage{hyperref}

\renewcommand{\d}{\mathrm{d}}

\newcommand{\be}{\begin{equation}}
\newcommand{\ee}{\end{equation}}
\newcommand{\bea}{\begin{eqnarray}}
\newcommand{\eea}{\end{eqnarray}}
\newcommand{\bse}{\begin{subequations}}
\newcommand{\ese}{\end{subequations}}

\newcommand{\pf}{k_{\mathrm F}}
\newcommand{\kf}{k_{\mathrm F}}
\newcommand{\vf}{v_{\mathrm F}}
\newcommand{\ef}{\varepsilon_{\mathrm F}}

\renewcommand{\d}{\mathrm{d}}

\newcommand{\rtf}{R_\mathrm{TF}}
\newcommand{\mathsym}[1]{{}}

\begin{document}

\title{Landau instability and mobility edges of the interacting one-dimensional Bose gas in weak random potentials}

\author{Alexander Yu.~Cherny}\email{cherny@theor.jinr.ru}
\affiliation{Bogoliubov Laboratory of Theoretical Physics, Joint Institute for Nuclear
Research, 141980, Dubna, Moscow region, Russia}

\author{Jean-S\'ebastien Caux}
\affiliation{Institute for Theoretical Physics,
Science Park 904,
University of Amsterdam, 1098 XH Amsterdam, The Netherlands}

\author{Joachim Brand}
\affiliation{Dodd-Walls Centre for Photonic and Quantum Technologies, Centre of Theoretical Chemistry and Physics, New Zealand Institute for Advanced Study,   Massey University, Auckland, New Zealand}

\date{\today}

\begin{abstract}
We study the frictional force exerted on the trapped, interacting 1D Bose gas under the influence of a moving random potential. Specifically we consider weak potentials generated by optical speckle patterns with finite correlation length. We show that repulsive interactions between bosons lead to a superfluid response and suppression of frictional force, which can inhibit the onset of Anderson localisation. We perform a quantitative analysis of the Landau instability based on the dynamic structure factor of the integrable Lieb-Liniger model and demonstrate the existence of effective mobility edges.
\end{abstract}


\maketitle

\section{Introduction}
Transport phenomena are behind many exciting developments in condensed matter physics from the discovery of superfluids and superconductors to the quantum Hall effect and topological insulators. In particular the features of superfluid flow have been studied intensely with ultra-cold atoms \cite{raman99:critical_velocity,Miller2007,Ramanathan2011}. On the other hand transport across random potentials has been used to verify the effect of Anderson localization, where interference from randomly distributed scatterers conspires to localize waves and thus inhibit transport \cite{Anderson1958}. Experimental tests of Anderson localization have been performed in Paris and Florence where a trapped one-dimensional Bose-Einstein condensate was expanded by dropping the trap in the presence of a speckle generated random potential \cite{billy08,roati08,deissler10}. In addition to the properties of wave propagation, transport measurements can also probe and reveal the nontrivial many-body nature of a quantum fluid. Of special interest are low-dimensional systems where strong correlations can be important and the manifestations of superfluidity are subtle \cite{Cherny2012}. Luckily, in one dimension exact solutions of the many-body problem are available and enable us to generate quantitative theoretical results for comparison with experiments.

While superfluidity is a collective effect of a many-body systems, the phenomenon of
Anderson localization is  a single-particle effect that affects linear waves in a random potential \cite{Anderson1958}.
The influence of interparticle interactions on the effect of Anderson localization is a long-standing problem and has been studied by many authors (see, e.g., Refs.~\cite{sanchez10:rev,modugno10:rev,aleiner10,shapiro12:rev,Ivanchenko2014,Flach2010,Larcher2012,brezinova12,Geiger12,Dujardin16} and references therein). Most of these studies consider the long-term effect of a random potential on allowing or prohibiting transport of ultra-cold atoms. Another interesting question concerns the role of superfluidity and the mechanism of its breakdown:
When a superfluid gas is subjected to a weak random potential, the property of superfluids to support frictionless flow may lead us to anticipate that the regime of Anderson localisation may never be reached (or only be reached at extremely long time scales).
In this work we consider the question of the breakdown of superfluid flow in the presence of a weak speckle potential based on exact results for the dynamic structure factor of the one-dimensional Bose gas.

Previous work on superfluidity of the 1D Bose gas has established that weak interactions indeed   make the system ``insensitive" to small external perturbation of arbitrary nature \cite{cherny09a,Cherny2012,kagan00}. On the other hand, increasing the strength of the interparticle interactions brings the gas into the Tonks-Girardeau regime, which is similar to a free Fermi gas and thus cannot be regarded as a universal superfluid.
The question arises about the mechanism  of the breaking of superfluidity in random fields, and its relation to the mobility edges of Anderson localization. In Anderson localization of linear waves, a mobility edge is an energy threshold that marks the transition between localized eigenstates inhibiting transport and extended eigenstates allowing transport \cite{Lee1985,Evers2008}. The mobility edge of a three-dimensional weakly interacting Bose gas was recently measured \cite{semeghini15} and calculations for laser speckle potentials for non-interacting atoms with the transfer-matrix method appeared in Ref.~\cite{Delande2014a}.

For independent particles moving in a random potential in one dimension there is no true mobility edge \cite{Lee1985,beenakker97}. However, as was shown \cite{sanchez07,lugan07a,gurevich09} for a random potential with a \emph{finite} correlation length $\sigma_\mathrm{r}$, the Lyapunov exponent is equal to zero for a plane wave spreading with the wavevector $k>1/\sigma_\mathrm{r}$. This implies the existence of a mobility edge  at the energy $E_\mathrm{mob} = \hbar^2/(2m\sigma_\mathrm{r}^2)$ for non-interacting particles. Hence, the dynamical transition to an Anderson localised state is suppressed when $k>1/\sigma_\mathrm{r}$. Strictly speaking this is true only for a weak random potential and finite time scales. Technically, the suppression arises at the level of the Born approximation, i.e.\ in the leading order term of a series expansion in powers of the dimensionless parameter $\epsilon_\mathrm{R}\equiv 2m \sigma_\mathrm{r}^2 V_\mathrm{R}/\hbar^2$ [here $V_\mathrm{R}$ is the mean amplitude of the random potential, see Eq.~(\ref{gk}) below] \cite{lugan09}. Taking into account the next terms in the Born series yields a series of sharp crossovers for the exponent, whose value drops at $k_n=n/\sigma_\mathrm{r}$  ($n=1,2,\ldots$) by orders of  magnitude. The smaller the amplitude of the random field, the larger suppression of  the Lyapunov exponent even for $k>1/\sigma_\mathrm{r}$. For the purpose of this work we consider small random field perturbations moving relative to an interacting one-dimensional Bose gas. The Born approximation is thus valid and the response of the superfluid can be evaluated from linear response theory. Effective mobility edges then arise from an interplay of the finite correlation length of the random potential and the superfluid response properties.

A link between superfluidity (a collective effect) and Anderson localization (a single-particle effect) is provided by the Landau criterion of superfluidity. It predicts uninhibited fluid motion relative to small-amplitude potentials of {arbitrary} shape at speeds slower than the critical velocity $v_\mathrm{c}$, which imposes a lower bound on the effective mobility edge: $E_\mathrm{mob}\geqslant m v_\mathrm{c}^2/2$. The critical velocity of a repulsive weakly-interacting Bose gas coincides with the speed of sound, which is proportional to the square root of the interaction strength. By contrast, the mobility edge of Anderson localization of non-interacting particles in a random potential with vanishing correlation length equals zero. Thus the Landau criterion here mandates an increase of the mobility edge proportional to the interaction strength.

In the case of the non-interacting Bose gas in a random potential with finite correlation length, however, the usual Landau criterion severely underestimates the mobility edge, since the Landau critical velocity is just zero. On the other hand,  a \emph{generalized} Landau criterion based on quantifying the drag force \cite{cherny09a,Cherny2012}, not only successfully reproduces the mobility edge for non-interacting particles (see the end of Sec.~\ref{an_res} below) but also applies to a system with arbitrary interparticle interactions moving in a weak random potential.

In this work we apply these ideas to a repulsively interacting one-dimensional Bose-gas of atoms in a moving weak laser speckle potential. Except for the moving random potential we consider the gas to be in equilibrium, e.g.\ contained in a time-independent trapping potential. Note that our approach does not strictly apply to the situation of an expanding Bose gas after trap release realised in experiments \cite{billy08,roati08} because the interacting one-dimensional Bose gas does not equilibrate locally during expansion and thus the assumptions of our approach do not apply in this case \cite{Campbell2015}. Instead we assume that only the speckle potential moves relative to the gas. Moving the speckle potential across the Bose gas at sufficiently high velocities, where superfluidity breaks down, will create excitations, which we quantify by calculating the mutual drag force based on linear response theory and the dynamic structure factor of the one-dimensional Bose gas \cite{cherny09}. The magnitude of the drag force thus provides a quantitative generalization of Landau's criterion of superfluidity \cite{cherny09a,Cherny2012} by  giving us the dissipation rate of the Landau instability. The main finding is that effective mobility edges emerge due to a combination of the finite momentum range of experimentally generated laser speckle \cite{modugno10:rev} and the characteristic shape of the dynamic structure factor (see the discussion
in Sec.~\ref{sec:linear} below).

The effective mobility edges separate the regime of zero drag force from that of finite drag force, and thus the separation line is interpreted as the dynamical onset of Anderson localization. When a finite drag force is present, the superfluid state of the Bose gas will eventually be destroyed and Anderson or many-body localisation phenomena will govern the evolution of the system for long times. While a finite drag force is a prerequisite for Anderson localization to develop, this approach cannot provide details about the statics or dynamics of Anderson or many-body localized phases, which can be obtained by other methods \cite{sanchez10:rev,modugno10:rev,aleiner10,shapiro12:rev}.

The absence of the drag force means that the system is superfluid and thus stable against external perturbations. Then the linear response theory is quite applicable at least in the vicinity of the onset of non-zero values of the drag force. This implies that effective mobility edges can be calculated with linear response theory.

For a weakly interacting Bose gas, only Bogoliubov's type of excitations is important ($\omega_+$ in Fig.~\ref{fig:dsf}). The speed of sound is then the critical velocity of superfluidity breakdown, which in conjunction with the density profile of the trapped gas cloud should provide an effective mobility edge. However, if the effective interaction constant $\gamma$ increases, \emph{subsonic} velocities generate drag as well. Nevertheless, frictionless flow may still persist at small velocities if the external perturbing potential has a limited momentum range, as is the case for laser-generated speckle. In this case, the Lieb type II
elementary excitations ($\omega_-$ in Fig.~\ref{fig:dsf}) provide a second, ``soft" mobility edge. The Landau instability takes place between the two mobility edges in the form of a continuous transition. In the limiting case of infinite $\gamma$ the transport behaviour of the 1D Bose gas is equivalent to that of the free Fermi gas (Tonks-Girardeau gas), because infinite contact repulsions emulate the Pauli principle.

The paper is organized as follows. After introducing the model in Section \ref{sec:model}, we outline the quantification of its rate of dissipation using linear response theory in Section \ref{sec:linear}. The disappearance of superfluidity and mobility edges are discussed in Section \ref{sec:superfluidity}, and a harmonically trapped gas in a moving random potential is discussion in Section \ref{sec:expansion}, followed by our conclusions.

\section{The model}
\label{sec:model}
Cold bosonic atoms confined to a waveguide is modeled by a one-dimensional gas of $N$
bosons with contact repulsive interactions (see, e.g., \cite{olshanii98,cherny04})
\begin{equation}
H =  \sum_{i=1}^N -\frac{\hbar^2}{2 m}\frac{\partial^2}{\partial x_i^2}
+ g_{\text{B}} \sum_{1\leqslant i<j\leqslant N} \delta(x_i - x_j)+\sum_{i=1}^N\frac{m\omega^2x_i^2}{2}.
\label{LLham}
\end{equation}
The last term is a harmonic potential with frequency $\omega$, trapping the system along the waveguide. In the absence of the trapping potential ($\omega=0$), this system of bosons is known as the Lieb-Liniger model \cite{lieb63:1}. The dimensionless parameter $\gamma\equiv m g_{\text{B}}/(\hbar^2 n)$ measures the strength of interactions, where $m$ and $n=N/L$ are the mass and the linear density, respectively. For infinitely strong repulsions $\gamma\to \infty$, the resulting model is known as the Tonks-Girardeau (TG) gas. In this limit the Bose gas can be mapped one-to-one to a non-interacting spinless {Fermi} gas, because infinite contact repulsions emulate the Pauli principle \cite{girardeau60}. The limit of small $\gamma$ corresponds to the well-known Bogoliubov model of weakly interacting bosons \cite{lieb63:1} (see also \cite{stringari96,yang14}).

\section{Linear Response to a Random Potential}
\label{sec:linear}
The rate of dissipation caused by a moving external perturbation (say, a point-like obstacle or random potential or shallow lattice) is connected to a local drag force, that is, momentum per unit time transferred to the gas from the external potential during motion. For the inhomogeneous Bose gas, one can apply the local density approximation if the density varies slowly on the length scale of a healing length \cite{kheruntsyan05}. The problem is then reduced to that of calculating the drag force in the homogeneous system.

The force can be calculated in the limit of small-amplitude external potential with the formalism of linear response theory \cite{astrakharchik04,cherny09a,Cherny2012,lang15}. It is convenient to choose the frame of reference where the gas is at rest but the external potential is moving with constant velocity $v$. This trick does not influence the resulting dissipation rate. The perturbation takes the form $\sum_j V_\mathrm{}(x_j-vt)$, where $V_\mathrm{}(x)$ is the energy of one boson in the stationary external perturbative potential, and the summation is over all the bosons. The dissipation rate is connected to the probabilities of transitions to excited states characterized by certain momentum and energy transfers. This probability is encoded in the dynamic structure factor (DSF), which relates to the time-dependent density correlator through Fourier transformation. It is given by the definition \cite{pitaevskii03:book}
\begin{equation}
S(k,\omega)={\cal Z}^{-1}\sum_{n,m}|\langle m|\delta\hat{\rho}_{k}|n \rangle|^{2}
 \delta(\hbar \omega-E_{n}+E_{m}),
\label{sqomega0}
\end{equation}
with ${\cal Z}=\sum_{m}\exp(-\beta E_{m})$ being the partition function and $\beta$ being the inverse temperature.
Here $\delta\hat{\rho}_{k}$ is the Fourier component of the density fluctuation, and $|n \rangle$ and $E_{n}$ are the $n$-th state and energy of the many-body system, respectively.

We obtain the value of drag force for the perturbation potential $V_\mathrm{}(x)$ \cite{cherny09a,Cherny2012}
\begin{equation}
F_{\mathrm{v}}=\int_{0}^{+\infty}\d k\,k|\widetilde{V}_\mathrm{}(k)|^2 S(k,k v)[1-\exp(-\beta \hbar k v)]/L
\label{dragf1D}
\end{equation}
with $\widetilde{V}_\mathrm{}(k)$ being the Fourier transform of the external potential $V_\mathrm{}(x)$.
This is the most general form of the drag force within linear response theory. At zero temperature, the second term in brackets is equal to zero.

Once the dynamic structure factor is known, the transport properties for any kind of potential can be calculated. Here we consider the special case of a speckle pattern generated from a diffusive plate that is illuminated by laser light.
In order to examine the transport properties of a random potential it is useful to consider an ensemble of individual realizations of potentials and later averaging over the ensemble. Potentials created from laser speckle are characterized by a correlation function $\langle V_\mathrm{}(x)V_\mathrm{}(x')\rangle=g(x - x')$, where $\langle\cdots\rangle$ stands for the ensemble average. The average properties of the drag force can be calculated by averaging the drag force of Eq.~(\ref{dragf1D}). We obtain
\begin{equation}
\langle F_\mathrm{v}\rangle  = \int_0^{2k_C} \d k\,k \widetilde{g}(k)S(k,k v)[1-\exp(-\beta \hbar k v)].
\label{dfrand}
\end{equation}
Here $\widetilde{g}(k)\equiv\langle |\widetilde{V}_\mathrm{}(k)|^2\rangle/L$ is the Fourier transform of the correlation function $g(x)$.
The integral limits in  (\ref{dfrand}) arise from the finite support of  the function $\widetilde{g}(k)$ originating in
the limited aperture of the diffusion plate generating the random phase \cite{goodman75:book,clement06,modugno10:rev}. Therefore, $\widetilde{g}(k) = 0$ for $|k| > 2k_C$. For estimations, we take a realistic correlation function \cite{goodman75:book,clement06,modugno10:rev}
\begin{equation}\label{gk}
\widetilde{g}(k) = \pi V_\mathrm{R} ^2\sigma_\mathrm{r}\mathrm{\Theta}\!\! \left( 1 - \frac{|k|\sigma_\mathrm{r}}{2}\right)\left( 1 - \frac{|k|\sigma_\mathrm{r}}{2}\right).
\end{equation}
Here $\Theta$ is the Heaviside step function, and $\sigma_\mathrm{r}\equiv 1/k_C$ is the random potential correlation length,
depending of the parameters of the experimental device, and $V_\mathrm{R}$ is the mean height of the barriers created by the laser beam. {Note that the correlation function $g(x)$ proportional to the $\delta$-function (the white-noise disorder) can be obtained in the limit $V_\mathrm{R} ^2\sigma_\mathrm{r} =\mathrm{const}$ and $\sigma_\mathrm{r} \to 0$.}

\begin{figure}[tb]
\centerline{\includegraphics[width=\columnwidth,clip=true]{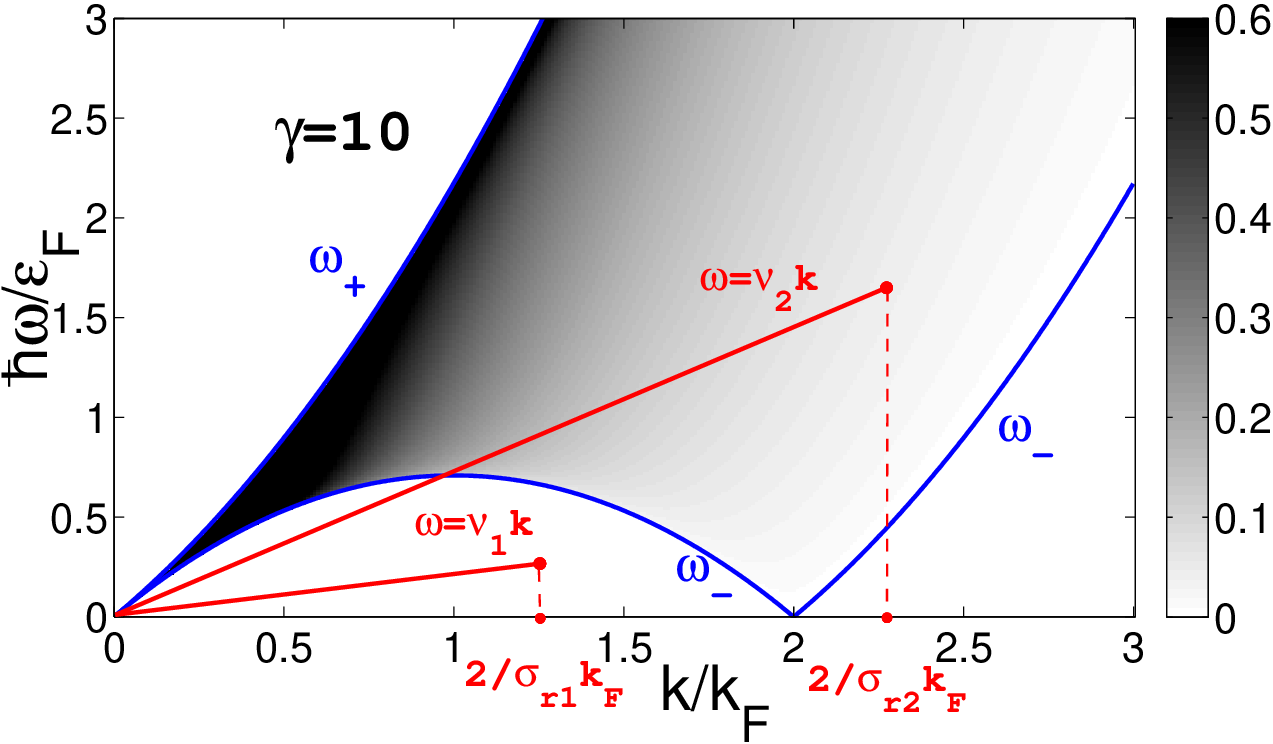}}
\caption{\label{fig:dsf} (Color online) Dynamic structure factor of the 1D Bose gas. The dimensionless value of the rescaled $S(\omega,k)\ef /N$ is coded in grey scale from ABACUS data. The Fermi energy $\ef=\hbar^2\pi^2 n^2/(2m)$ is used as a unit of energy. The straight (red) lines show the path of integration for evaluating the frictional force in Eq.~(\ref{dfrand}) for two different values of the relative velocity $(v_1, v_2)$ between the Bose gas and speckle potential with correlation lengths $\sigma_{\mathrm{r}1}$ and $\sigma_{\mathrm{r}2}$, respectively. The usual Landau criterion stops working here, since some excitation states always lie below a straight line with arbitrary slope; this is due to the presence of umklapp excitations $\omega_{-}(2\kf)=0$.
}
\end{figure}

In order to calculate the drag force, we need to know the DSF of the Lieb-Liniger model, given by the Hamiltonian (\ref{LLham}). The exact integrability of the Lieb-Liniger model now permits the direct numerical calculation of dynamical correlation functions such as the DSF \cite{caux06} for systems with finite numbers of particles by means of the algebraic Bethe ansatz \cite{korepin93:book} using the ABACUS
algorithm \cite{2009_Caux_JMP_50}.
Another way to evaluate the DSF is to use a simple interpolating expression \cite{cherny09}, whose values deviate from the ABACUS calculations within a few percent \cite{cherny09}. The generic behaviour of the DSF is shown in Fig.~\ref{fig:dsf}.

For the TG gas, the DSF is given in the thermodynamic limit by
\begin{equation}
S(k,\omega)\frac{\ef}{N}= \frac{\kf}{4 k}
\label{DSFTG}
\end{equation}
for $\omega_{-}(k)\leqslant\omega\leqslant\omega_{+}(k)$, and zero otherwise \cite{brand05,cherny06}. Here $\omega_\pm(k)$ are the energy dispersions bounding a single quasiparticle-quasihole excitation.
The branches  $\omega_{+}$ and $\omega_{-}$ correspond to the Lieb's type I and II excitations, respectively \cite{lieb63:2}. They are known analytically in the TG regime
\begin{equation}\label{omplmi}
\omega_\pm(k)={\hbar |2 \pf k \pm k^2|}/{(2 m)},
\end{equation}
where we have used
 $\kf\equiv\pi n$ and $\ef\equiv\hbar^{2}\kf^{2}/(2m)$ for the Fermi wave vector and energy of the TG gas, respectively. The sound velocity is given by $\mathrm{d}\omega_\pm(k)/\mathrm{d}k $ at $k=0$ and equal to $\vf=\hbar\kf/m$.

In the Bogoliubov regime of small interactions $\gamma\ll 1$, we have \cite{pitaevskii03:book}
\begin{equation}
S(k,\omega)=N\frac{T_{k}}{\hbar\omega_k} \delta(\omega-\omega_{k}),
\label{dsfbog}
\end{equation}
where $T_{k}=\hbar^{2}k^{2}/(2m)$ and
\begin{equation}\label{ombog}
\hbar\omega_k=\sqrt{T_k^{2}+2n g_{\text{B}} T_k}
\end{equation}
are the free-particle and Bogoliubov energy spectrum, respectively. For small but finite values of $\gamma$, the upper branch $\omega_+(k)$ remains very close to the Bogoliubov energy spectrum \cite{lieb63:2}, and non-zero values of the DSF are located near this branch thus emulating the $\delta$-function behaviour of the DSF \cite{brand05,cherny06}. The sound velocity in the Bogoliubov regime is equal to $\sqrt{n g_{\text{B}}/m}$.

\begin{figure}[tb]
\centerline{\includegraphics[width=.96\columnwidth]{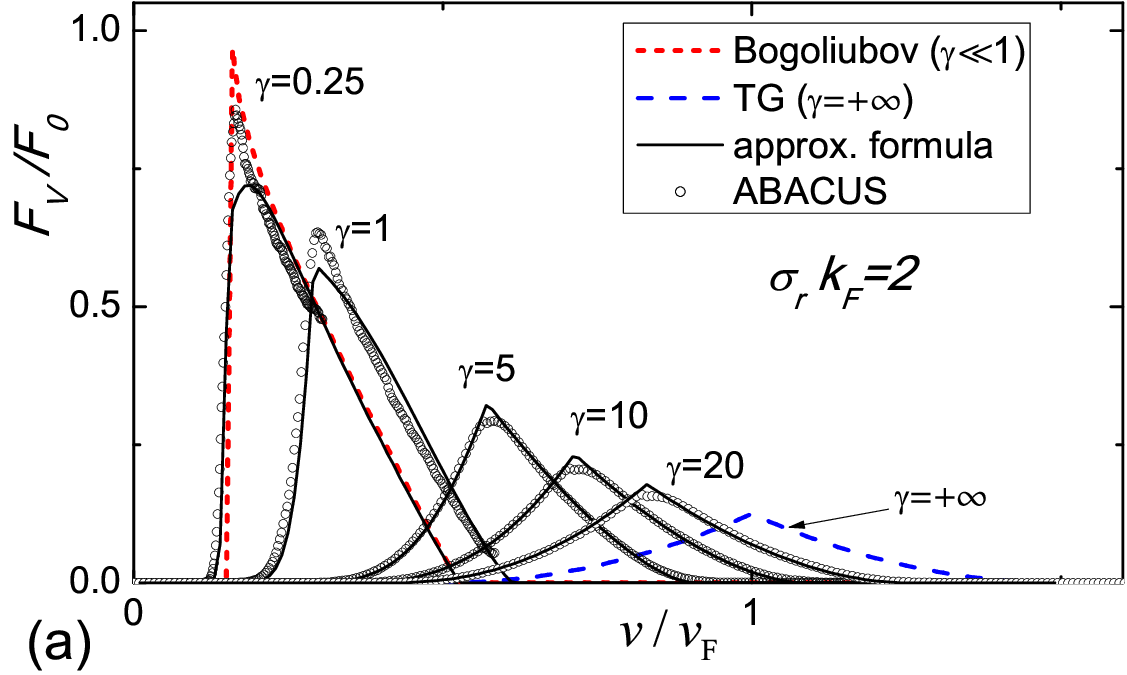}}
\centerline{\ \includegraphics[width=.98\columnwidth]{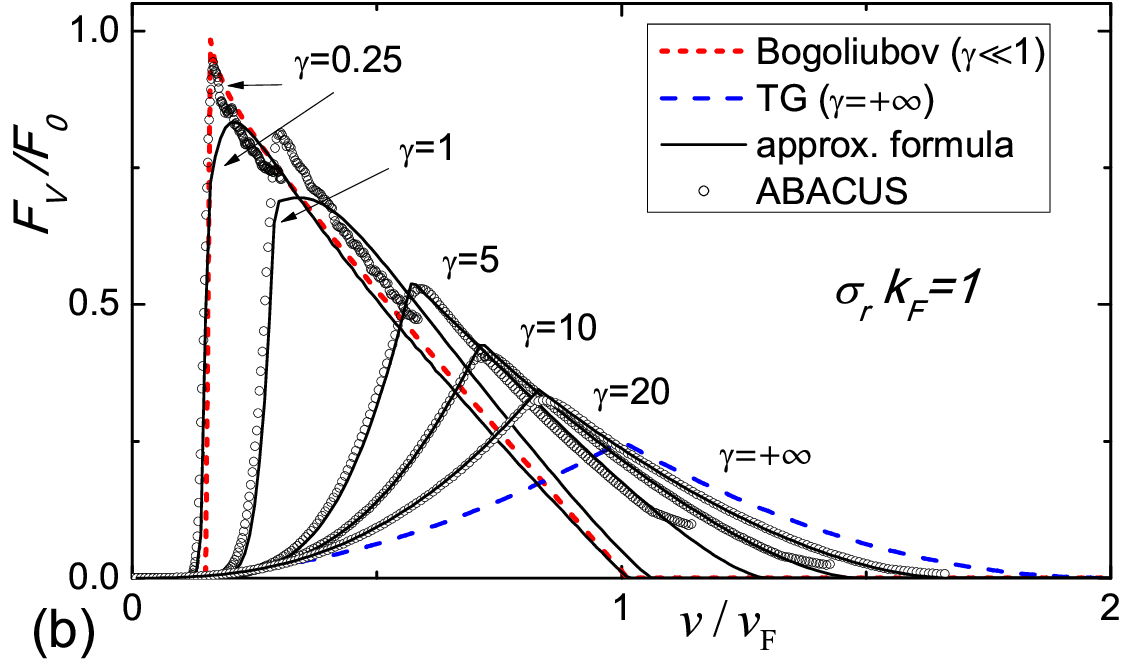}}
\centerline{\includegraphics[width=.98\columnwidth]{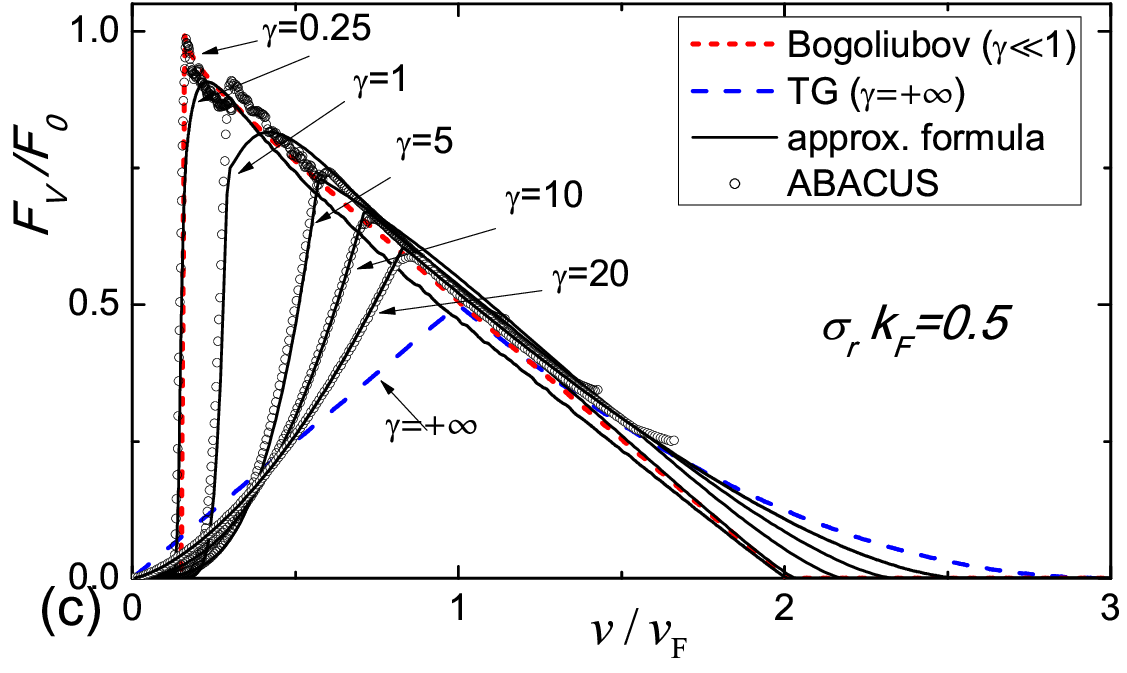}}
\caption{\label{fig:dfrandom} (Color online) Frictional force (in units of $F_0\equiv{2\pi m V_\mathrm{R} ^2\sigma _r N}/{\hbar ^2}$) as a function of velocity (in units of $\vf\equiv\hbar \pi n /m$) for different values of the interaction parameter $\gamma$ and the random potential correlation length $\sigma_\mathrm{r}$ [in units of $1/\kf=1/(\pi n$)]. Numerical ABACUS data [from DSF results extrapolated to infinite system size from finite particle number exact solutions of the Lieb-Liniger model \eqref{LLham}] is compared to the approximate expression from Ref.~\cite{cherny09} and the limiting expressions for the small and large $\gamma$ cases.
}
\end{figure}

Integration of the dynamic structure factor over the lines indicated in Fig.~\ref{fig:dsf} yields the frictional force in accordance with Eq.~(\ref{dfrand}). The control parameters governing the drag force are the potential velocity, the interaction strength, and the correlation length. The results are depicted in Fig.~\ref{fig:dfrandom}. The ABACUS data are obtained for $N=300$ particles at $\gamma=0.25$, $N=200$ ($\gamma=1$), and $N=150$ ($\gamma=5;10;20$).

The interpolation formula works well at subsonic and supersonic velocities, but is slightly worse in the vicinity of sound velocity. In contrast, due to incomplete saturation of the sum rule at high momentum, the ABACUS overrates the values of the force at sufficiently large velocities. At small $\gamma$ (Bogoliubov regime), in order to compute the force at sufficiently large velocities we must significantly increase the number of particles in the ABACUS calculations, but this is not needed since this region is extremely well described by the interpolation formula.

\section{Disappearance of Superfluidity and Mobility Edges}
\label{sec:superfluidity}
In one dimension, there is no \emph{qualitative} criterion for superfluidity due to the absence of the long-range order; however, one can suggest a \emph{quantitative} criterion \cite{cherny09a,Cherny2012}. The value of the drag force can  be  used to map out a zero-temperature phase diagram for the superfluid--insulator transition: superfluidity assumes zero or strongly suppressed values of the drag force.
This criterion can be quite effective in practice.
For instance, even for quite moderate value of the coupling parameter $\gamma=0.25$, the drag force for subsonic and supersonic velocities can differ by 45 orders of magnitude \cite{Cherny2012}!

All the results shown in Fig.~\ref{fig:dfrandom} can easily be understood with Eq.~(\ref{dfrand}) and the $k$-$\omega$ diagram of Fig.~\ref{fig:dsf}. Changing the velocity $v$ of moving potential leads to rotating the segment of integration about the origin of the coordinates in the $k$-$\omega$ plane. The length of the segment is determined by the correlation length $\sigma_\mathrm{r}$ and density $n$. The value of frictional force is close to zero at small and large velocities, since the
DSF vanishes almost everywhere along the segment of integration. For instance, if $\sigma_\mathrm{r}> 1/(\pi n)$ then the drag force vanishes exactly at sufficiently small velocities, because the DSF equals to zero below Lieb's type II dispersion due to the conservation of both energy and momentum \cite{lieb63:1}. The borders of localization of the drag force in velocity space can be calculated analytically in the Bogoliubov and TG regimes (see the next section below). The drag force reaches its maximum at sound velocity, since the DSF takes non-zero values at small momenta along the segment of integration.

\subsection{Analytical results for the drag force in the Bogoliubov and Tonks-Girardeau limits}
\label{an_res}

As shown above, the DSF is known analytically in the Bogoliubov and TG regimes, which enables us to calculate the drag force analytically, following \cite{Cherny2012}. For small values of $\gamma$, we obtain from Eqs.~(\ref{dfrand}) and (\ref{dsfbog})
\begin{equation}
\langle F_{\mathrm{v}}\rangle  = F_0\,\Theta(\tilde{v}-\tilde{v}_c)\Theta(1-z)(1-z).
\label{dfrandBog}
\end{equation}
Here  $F_0\equiv{2\pi m V_\mathrm{R} ^2\sigma _r N}/{\hbar ^2}$ is a unit of force, $z\equiv\pi n \sigma_\mathrm{r}\sqrt{\tilde{v}^2-\tilde{v}_c^2}$, $\tilde{v}\equiv v/\vf$ and $\tilde{v}_c=\sqrt{\gamma}/\pi$ are the velocity and sound velocity, respectively, in units of $v_\mathrm{F}$.
In the TG regime, Eqs.~(\ref{DSFTG}) and (\ref{dfrand}) yield
\begin{equation}
\langle F_{\mathrm{v}}\rangle  = F_0[f_1+(f_2-f_1)\Theta(\lambda_+ -\lambda_0)-f_2\Theta(\lambda_- -\lambda_0)],
\label{dfrandTG}
\end{equation}
where we introduce the notations $\lambda_0 \equiv 2/(\pi n \sigma_\mathrm{r})$, $\lambda_\pm\equiv2|\tilde{v}\pm 1|$, $f_1\equiv\frac{1}{4}(\lambda_+-\lambda_-)(1-\frac{\lambda_+ +\lambda_-}{2\lambda_0})$, $f_2\equiv\frac{(\lambda_0-\lambda_-)^2}{8\lambda_0}$. The expressions (\ref{dfrandBog}) and (\ref{dfrandTG}) first appeared in Ref.\ \cite{Cherny2012}.

\begin{figure}[tb]
\centerline{\includegraphics[width=.95\columnwidth]{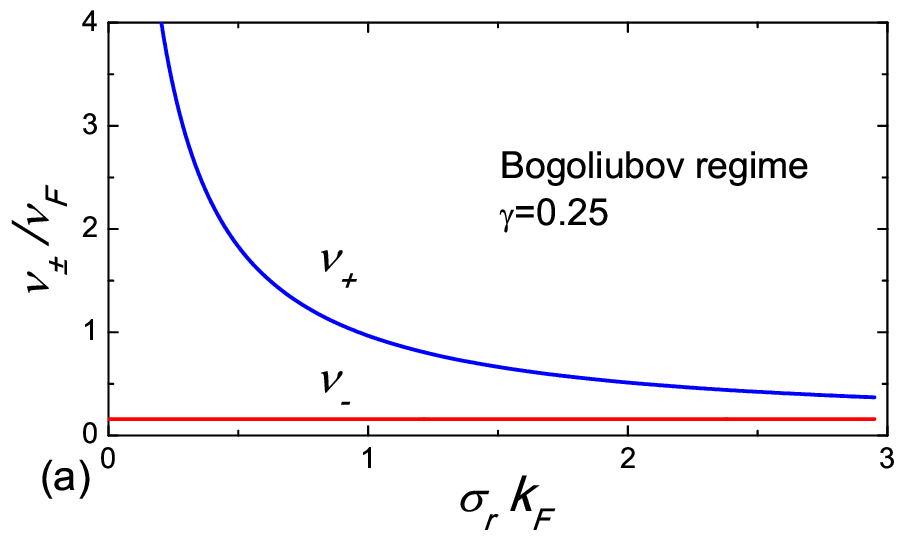}}
\centerline{\ \includegraphics[width=.95\columnwidth]{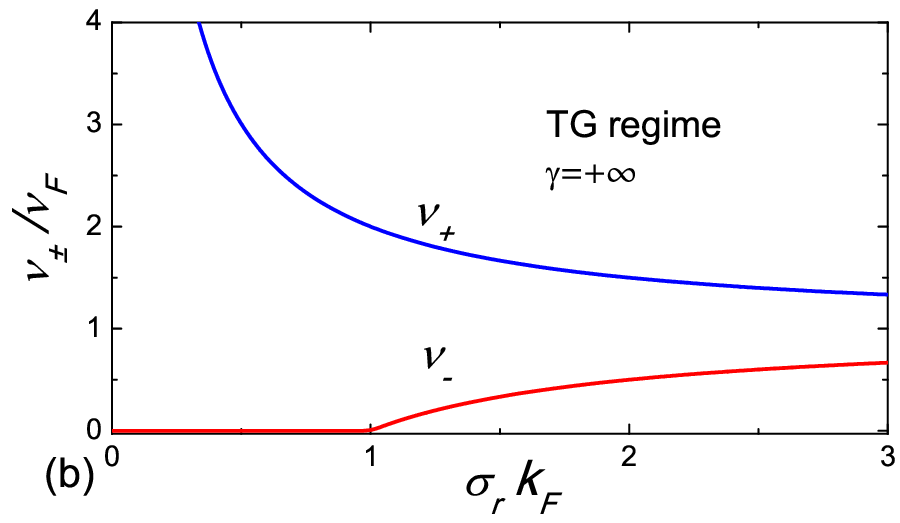}}
\caption{\label{fig:vpm} (Color online) The borders of localization of the drag force in the Bogoliubov (a) [Eqs.~(\ref{vminus_bog}), (\ref{vplus_bog})] and TG (b) [Eqs.~(\ref{vminus_TG}), (\ref{vplus_TG})] regimes versus the random potential correlation length. The border velocities $v_{-}$ and $v_{+}$ and the correlation length $\sigma_\mathrm{r}$ are shown in units of $\vf\equiv\hbar \pi n /m$ and $1/\kf=1/(\pi n$), respectively. The results match  the drag force behaviour in the Bogoliubov and TG regimes represented in Fig.~\ref{fig:dfrandom}.}
\end{figure}

Having the analytical expressions for the drag force at our disposal, it is possible to determine at which velocities the drag force takes non-zero values. However, a  simpler way to find the borders of localization of the drag force is to use the $\omega$-$k$ diagram shown in Fig.~\ref{fig:dsf}. Indeed, the DSF is localized only along the upper branch $\omega_{+}(k)$ in the Bogoliubov regime, and the segment of integration in Eq.~(\ref{dfrand}) intersects the upper branch only above the sound velocity and below $\omega_{+}(2k_C)/(2k_C)$. Then the drag force is non-zero in the Bogoliubov regime only at velocities lying between $v_{-}$ and $v_{+}$, given by
\begin{align}
  v_{-}&=  \vf\frac{\sqrt{\gamma}}{\pi},\label{vminus_bog}\\
  v_{+}& = \vf\frac{1}{\kf\sigma_\mathrm{r}}\sqrt{1+(\kf\sigma_\mathrm{r})^2\frac{\gamma}{\pi^2}}.  \label{vplus_bog}
\end{align}
In the TG regime, the DSF is localized between $\omega_{-}(k)$ and $\omega_{+}(k)$, given by Eq.~(\ref{omplmi}). If $k_C>\kf$, the segment of integration always intersects $\omega_{-}(k)$ at sufficiently small velocities, and then the lower border $v_{-}$ is zero. Otherwise, if $k_C<\kf$, the lower border of the velocity range equals $\omega_{-}(2k_C)/(2k_C)$. The upper border is always given by the condition $\omega_{+}(2k_C)/(2k_C)$. By substituting Eq.~(\ref{omplmi}), we obtain the borders of localization of the drag force in the TG regime
\begin{align}
  v_{-}&= \left\{\begin{array}{ll}
                 0,& \kf\sigma_\mathrm{r}<1,\\[1ex]
                 \vf\big(1-(\kf\sigma_\mathrm{r})^{-1}\big),& \kf\sigma_\mathrm{r} \geqslant 1,
                   \end{array}\right.\label{vminus_TG}\\
 v_{+}& = \vf\big(1+(\kf\sigma_\mathrm{r})^{-1}\big).  \label{vplus_TG}
\end{align}
These results, shown in Fig.~\ref{fig:vpm}, are consistent with the behaviour of the drag force in the Bogoliubov and TG regimes represented in Fig.~\ref{fig:dfrandom}.

The maximum of DF in the TG regime is reached at the Fermi velocity, which can be seen from Figs.~\ref{fig:dsf} and \ref{fig:dfrandom}. After little algebra, Eq.~ (\ref{dfrandTG}) yields for $v=\vf$
\begin{align}
 F_{\mathrm{v}}^{\mathrm{(max)}}/F_0=\left\{\begin{array}{ll}
                 1-\kf\sigma_\mathrm{r},& \mathrm{if}\ 0<\kf\sigma_\mathrm{r}\leqslant 1/2,\\[1ex]
                 1/(4\kf\sigma_\mathrm{r}),& \mathrm{if}\ \kf\sigma_\mathrm{r} > 1/2.
                   \end{array}\right.
  \label{dfrandTGmax}
\end{align}

We emphasize that linear response theory yields the frictional force (\ref{dfrand}) for all values of interparticle interactions. The problem can be reduced effectively to the {one-particle problem} in a random potential in two limiting cases, the TG regime and, under a certain condition, the Bogoliubov regime.
It is well known that non-interacting particles experience a mobility edge if they move in a  random potential with the finite correlation length $\sigma_\mathrm{r}$. In this case, the mobility edge is given by \cite{sanchez07,lugan07a,gurevich09} $E_\mathrm{mob} = \hbar^2k_\mathrm{mob}^2/(2m)$ with $k_\mathrm{mob}=k_C\equiv1/\sigma_\mathrm{r}$. If $|k|>k_C$ the waves can propagate,  while in the opposite case the particle wave function is localized (Anderson localization), and transport is suppressed. In terms of velocities, the condition $v<\hbar k_C/m$ for the moving particle implies that it cannot move freely but is ``caught" by the random potential.

Let us point out that the results for the drag force in the TG regime are compatible with the existence of mobility edges for free particles. An argument based on the equivalence of the TG gas with free fermions can be found in Ref.\ \cite{Cherny2012}.

In the Bogoliubov regime of small $\gamma$, there is a characteristic length of the system called the healing length $\tilde{v}_\mathrm{h} \equiv \hbar/\sqrt{2\mu m}$, where $\mu$ is the chemical potential (see, e.g., Ref.~\cite{pitaevskii03:book}). In one dimension, the chemical potential is given by $\mu=g_{\text{B}} n$, and, hence, $\tilde{v}_\mathrm{h}=\pi/(\sqrt{2\gamma}\kf)$. Then in the regime $\tilde{v}_\mathrm{h}\ll\sigma_\mathrm{r}$, {the many-body effects dominate}. It follows from Eq.~(\ref{vplus_bog}) that $v_{+}$ is getting very close to $v_{-}$, and the system is superfluid at almost arbitrary velocities except for the close vicinity of the sound velocity. In the regime $\tilde{v}_\mathrm{h}\gg\sigma_\mathrm{r}$, the many-body effects are suppressed, and {the bosons behave as independent particles}. In this regime, $v_{+}\simeq \vf/(\kf\sigma_\mathrm{r})$, and there is no resistant force if $v>v_{+}$. This condition coincides with the condition of one-particle propagation $v>\hbar k_C/m$.

\subsection{A sum rule for the drag force}
\label{sumruleDF}

The drag force obeys a sum rule, which follows from the well-known $f$-sum rule for the DSF \cite{pitaevskii03:book}:
\begin{equation}\label{fsum}
\int_{0}^{+\infty}\d\omega\, \omega S(k,\omega)[1-\exp(-\beta \hbar \omega)]= N{k^{2}}/{(2m)}.
\end{equation}
The sum rule for the drag force can be obtained from Eq.~(\ref{dragf1D}) by multiplying it by $v$ and integrating over the velocity from zero to infinity. Making the substitution $v=\omega/k$ and using the f-sum rule (\ref{fsum}), we derive
\begin{equation}\label{DFsumrule}
\int_{0}^{+\infty}\d v\, v F_\mathrm{v}(v)= \frac{n}{2m}\int_{0}^{+\infty}\d k\,k |\widetilde{V}_\mathrm{}(k)|^2.
\end{equation}
This is the general form of the sum rule for the drag force, which is valid for an arbitrary external potential. The right-hand-side of the sum rule is {independent of interactions between particles and temperature}. Note that $v F_\mathrm{v}$ is nothing else but \emph{the rate of energy dissipation}, that is, the energy loss per unit time in the reference frame where the system is at rest but the potential moves with velocity $v$.

In order to specify the sum rule for a random potential, we need to take the average of Eq.~(\ref{DFsumrule}) over the random potential ensemble as we did while deriving Eq.~(\ref{dfrand}). In this manner, we obtain with the specific form of the correlation function of the random potential given by Eq.~(\ref{gk})
\begin{equation}\label{DFsumrulerand}
\int_{0}^{+\infty}\d v\, v \langle F_\mathrm{v}(v)\rangle= \frac{N}{3m}\frac{\pi V_\mathrm{R}^2}{\sigma_\mathrm{r}}.
\end{equation}

In this paper, the drag force is used as a quantitative measure of superfluity in one dimension. From this point of view, we arrive at the seemingly paradoxical conclusion with the sum rule (\ref{DFsumrulerand}) for the drag force that interactions, in a way, {do not influence superfluidity}. Indeed, though the value of drag force depends on the strength of interactions at a given velocity, its ``integral value" given by the left-hand-side of Eq.~(\ref{DFsumrulerand}) does not. Moreover, it also depends on neither temperature nor the type of statistics. The latter follows form the fact that the sum rules (\ref{DFsumrule}) and (\ref{DFsumrulerand}) are obtained in a very general way without using the bosonic or fermionic nature of the system. Thus, the ``integral value" of the energy dissipation rate is independent of interactions not only for random  but for {arbitrary} potentials.
All these contributing factors (interactions, statistics, temperature, details of perturbing potential) do, of course, influence the velocity-dependent dissipation rate, as seen in the previous sections. The sum rule is valid within the linear response method, which is, in effect, the time-dependent perturbation theory of the first order. Beyond the linear response regime, the integral value  is changed.

\section{A harmonically trapped gas in a moving random potential}

The results of the previous section, shown in Fig.~\ref{fig:dfrandom}, enable us to understand an experimentally more reliable case of the trapped 1D Bose gas in a random field, moving with constant velocity $v$.

The density profile of the gas, described by Eq.~(\ref{LLham}), can be determined from the equation of state via the LDA (Thomas-Fermi) approximation. Then the drag force in the linear response formalism is written as an integral over local contributions, i.e. where the Bose gas can be assumed to be in local equilibrium and well described by the LDA. We explicitly consider the case of strong interactions $\gamma \gg 1$ (the TG gas), where simple closed-form expressions are found.

\subsection{The density profile of the TG gas}
\label{nvTG}
Let us consider the TG gas of $N$ atoms, trapped by a 1D harmonic potential with frequency $\omega$, in the local density approximation. Since the TG gas can be mapped exactly into the Fermi gas \cite{cheon99}, the local density approximation for the system is nothing else but the well-known Thomas-Fermi approximation (see, e.g., Ref.~\cite{pitaevskii03:book,kheruntsyan05}). Within the approximation, the initial profile of the density at $t=0$ is given by
\begin{equation}
n(x)=n_0\sqrt{1-\frac{x^2}{\rtf^2}},\quad |x|\leqslant\rtf,
\label{nz}
\end{equation}
where
\begin{equation}\label{rtf}
\rtf=\frac{\hbar\pi n_0}{m\omega}
\end{equation}
is the Thomas-Fermi radius. The initial density in the center $n_0$ is related to the total number of particles and the frequency of the trapping potential by the formula
\begin{equation}
n_0=\left(\frac{2m\omega N}{\hbar}\right)^{1/2} \frac{1}{\pi}.
\label{n0}
\end{equation}
Thus, for describing the gas, we need to know two independent parameters $N$ and $\omega$. One can also use the frequency $\omega$ and the Thomas-Fermi radius $\rtf=\sqrt{2 N\hbar/(m\omega)}$ as independent control parameters.

\subsection{Drag Force in the local density approximation}
\label{sec:expansion}

In order to calculate the drag force in the local density approximation, one can use Eq.~(\ref{dfrand}) with the {local} parameters. It is convenient to measure the wave vector and frequency in the Fermi wave vector $\kf=\pi n$ and frequency $\ef/\hbar$, respectively. Thus we introduce \cite{Cherny2012} the dimensionless DSF $s(\lambda,\nu)\equiv\ef {S(\kf\lambda,\ef\nu/\hbar)}/{N}$, which is controlled in general only by the Lieb-Liniger parameter $\gamma$.

Within the local density approximation, the local Lieb-Liniger parameter and Fermi momentum are given by
\begin{equation}
\gamma=m g_{\text{B}}/[\hbar^2 n(x)],\quad k_\mathrm{Fl}=\pi n(x), \label{gam}
\end{equation}
respectively, where $n(x)$ is described by Eq.~(\ref{nz}). Then the drag force (\ref{dfrand}) \textit{per unit particle} takes the form
\begin{equation}
\frac{\langle F_{\mathrm{v}}(x)\rangle}{N}= f_0\int_{0}^{2/k_\mathrm{Fl}\sigma_\mathrm{r}}\d \lambda\
s(\lambda,2\lambda\tilde{v})\Big(1-\frac{k_\mathrm{Fl}\sigma_\mathrm{r}\lambda}{2}\Big),
\label{dfl}
\end{equation}
where $\tilde{v}$ is the velocity of moving random potential in units of the local Fermi velocity $v_\mathrm{Fl}(x)=\hbar\pi n(x)/m$. The unit of drag force is $f_0\equiv 2\pi m V_\mathrm{R}^2\sigma_\mathrm{r}/\hbar^2$. We emphasize that Eq.~(\ref{dfl}) is the local density approximation for the drag force, applicable in general. The coordinate dependence appears through the local velocity $\tilde{v}(x)$, Fermi wave vector $k_\mathrm{Fl}(x)$, and the Lieb-Liniger parameter $\gamma(x)$.

In the specific case of the TG gas at zero temperature, considered in the previous subsection, the DSF (\ref{DSFTG}) can be rewritten in the dimensionless variables
\begin{equation}\label{DSFTGdimless}
s(\lambda,\nu) = \frac{1}{4\lambda}[\Theta(\nu-\nu_{+})-\Theta(\nu-\nu_{-})],
\end{equation}
where $\nu_\pm=\lambda|\lambda\pm2|$. It follows from Eq.~(\ref{nz}) that the dimensionless velocity is given by
\begin{equation}
\tilde{v}\equiv \frac{v}{v_\mathrm{Fl}(x)}=\frac{v}{\omega\rtf}\frac{1}{\sqrt{1-{x^2}/{\rtf^2}}}.
\label{svf}
\end{equation}
Substituting  Eqs.~(\ref{DSFTGdimless}) and (\ref{svf}) into Eq.~(\ref{dfl}) yields the analytic expression
\begin{align}
  &\frac{\langle F_{\mathrm{v}}(x)\rangle}{N} =f_0\times \nonumber\\
  &\! \left\{\begin{array}{lll}
  0,& \!\mathrm{if}\ \frac{1}{\sigma}\leqslant |\tilde{v}_{0}-\alpha|,\\[1ex]
  \frac{\sigma}{4\alpha}\big[\frac{1}{\sigma}-|\tilde{v}_{0}-\alpha|\big]^2, & \!\mathrm{if}\ |\tilde{v}_{0}-\alpha|\leqslant\frac{1}{\sigma}\leqslant \tilde{v}_{0}+\alpha,\\[1ex]
  \frac{\tilde{v}_{0}}{\alpha}\Theta(\alpha\!-\!\tilde{v}_{0})\!+\!\Theta(\tilde{v}_{0}\!-\!\alpha)\!-\!\sigma\tilde{v}_{0}, & \!\mathrm{if}\ \frac{1}{\sigma}\geqslant \tilde{v}_{0}+\alpha,
  \end{array}\right. \label{df_TG}
\end{align}
where we put by definition $\sigma\equiv\sigma_\mathrm{r}\pi n_0 = \sigma_\mathrm{r}m\omega\rtf/\hbar$, $\tilde{v}_0\equiv \tilde{v}(0)={v}/({\omega\rtf})$, and $\alpha\equiv \sqrt{1-{x^2}/{\rtf^2}}$. The DF for the inhomogeneous TG gas, given by Eq.~(\ref{df_TG}), coincides with that of the homogeneous gas (\ref{dfrandTG}) when $x=0$ and $n_0=n$.

The first condition in Eq.~(\ref{df_TG})
\begin{equation}\label{SFcond}
\frac{\hbar}{\sigma_\mathrm{r}m\omega\rtf}\leqslant \left|\frac{v}{\omega\rtf}-\sqrt{1-\frac{x^2}{\rtf^2}}\right|
\end{equation}
 is actually the condition of superfluidity, discussed in detail in Sec.~\ref{an_res}. Note that if the velcity of random potential is sufficiently large then the drag force is zero for arbitrary point of the trapped gas. The DF reaches its maximum when the local velocity of sound (given by the local Fermi velocity in the TG regime) is equal to the velocity of the moving random potential
\begin{equation}\label{DFmax}
{v}={\omega\rtf}\sqrt{1-{x_{\mathrm{max}}^2}/{\rtf^2}}.
\end{equation}

It follows from the equations (\ref{SFcond}) and (\ref{DFmax}) that the edges of the superfluid regime in the trapped TG gas $x_{-}$ and $x_{+}$  and the point  where the DF attained its maximum $x_{\mathrm{max}}$ are given by
\begin{align}
  {x_{\mathrm{max}}}&=  \pm{\rtf}\sqrt{1-{\tilde{v}_0^2}},\label{zminus_TG}\\
 {x_{+}}& = \pm{\rtf}\sqrt{1-({\tilde{v}_0}-\sigma^{-1})^2},  \label{zplus_TG}\\
 {x_{-}}& = \pm{\rtf}\sqrt{1-({\tilde{v}_0}+\sigma^{-1})^2}, \label{zmax_TG}
\end{align}
where the velocity of the moving random potential is assumed to be positive. If the coordinates given by Eqs.~(\ref{zminus_TG})-(\ref{zmax_TG}) take complex values then the corresponding points lie beyond the TG localization $-\rtf\leqslant x\leqslant\rtf$.

\begin{figure}[tb]
\includegraphics[width=\columnwidth]{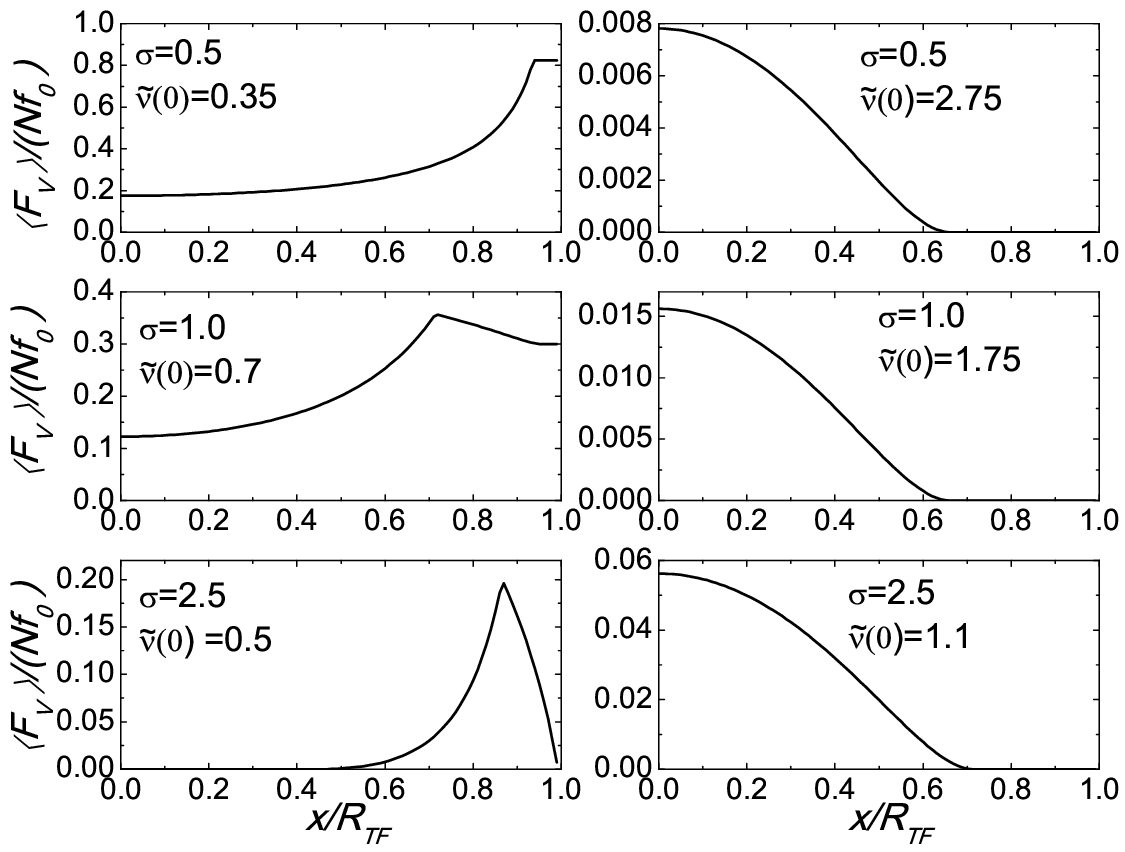}
\caption{\label{fig:DFtrapped}
The drag force per unit particle ($\langle F_{\mathrm{v}}\rangle/N$, in units of $f_0\equiv 2\pi m V_\mathrm{R}^2\sigma_\mathrm{r}/\hbar^2$) versus the coordinate ($x$, in units of the Thomas-Fermi radius $\rtf$) for a harmonically trapped Tonks-Girardeau gas, see Eq.~(\ref{df_TG}). The control parameters are the velocity of moving random potential in units of the sound velocity in the center [$\tilde{v}(0)\equiv v/(\omega\rtf)$] and the potential correlation length in units of the inverse Fermi wave vector in the center ($\sigma\equiv\sigma_\mathrm{r}m\omega\rtf/\hbar$). The frictional force is strictly zero below the lower edge $x_{-}$ and above the upper edge $x_{+}$, while it peaks in the point of coincidence $x_{\mathrm{max}}$ between the velocity of the moving potential and the local velocity of sound [see Eqs.~(\ref{zminus_TG})-(\ref{zmax_TG})]. It shows that the suppression of superfluidity is most noticeable near the velocity of sound.}
\end{figure}

The results for various values of the contrast parameters are shown in Fig.~\ref{fig:DFtrapped}.

\section{Conclusion}
\label{sec:conclusion}
In this paper, we have approached a problem of non-equilibrium quantum many-body dynamics from the perspective of integrable models. Starting from the recently improved understanding of the dynamic correlations of the one-dimensional Bose gas, it was possible to make quantitative predictions for non-trivial transport properties, which could be tested experimentally. Being based on exact results for the interacting quantum many-body system, our predictions go beyond the commonly employed mean-field approximations and nonlinear-wave models. In particular, we obtained the sum rule (\ref{DFsumrulerand}) for the drag force, which implies that interparticle interactions, in a way, do not influence the integrated drag force for a weak random potential at all (see the discussion in Sec.~\ref{sumruleDF}).

A severe limitation of our approach, however, stems from the use of linear-response theory, which is actually the first-order of the time-dependent perturbation theory. In the paper \cite{giamarchi87}, the a renormalization group method was applied to study superfluidity of the 1D Bose gas, which means that the contribution of the next orders of the perturbation theory were taken into consideration but only in the low-energy regime of the Luttinger liquid theory and for the random potential with zero correalation length. Thus, the usefulness of our results is restricted to weak random potentials but for the entire range of excitations in the $\omega-k$ plane, see Fig.~\ref{fig:dsf}. The severity of this limitation is difficult to evaluate, in particular, the conclusion about superfluidity of the 1D Bose gas at sufficiently large velocities provided the correlation length of the moving random potential are finite. It may require careful comparison with experimental data or possibly with fully quantum-dynamical simulations \cite{Ivanchenko2014} to answer this question.

\acknowledgments

The authors thank Sergej Flach, Peter Drummond, and Igor Aleiner for insightful discussion. A.~Yu.~Ch.\ acknowledges support from the JINR--IFIN-HH projects.
J.-S.~C.\ acknowledges support from the FOM and NWO foundations of the Netherlands. J.~B. received funding from the Marsden Fund of New Zealand (contract number MAU1604).

\bibliography{Drag_Force}

\end{document}